\documentclass[aps,prb,preprint,superscriptaddress]{revtex4-2}
\usepackage{amsmath,amssymb}
\usepackage{graphicx}  
\usepackage{dcolumn}   
\usepackage{bm}        
 \usepackage{float}
\usepackage{amsfonts}
\usepackage{gensymb}
\usepackage{upgreek}
\usepackage{mathrsfs} 
 \usepackage{hyperref}
\usepackage{epstopdf}
\usepackage{animate}

\hypersetup{
    colorlinks=true,
    linkcolor=blue,
    filecolor=magenta,      
    urlcolor=blue,
}
 
\usepackage{blindtext,tikz}
\usetikzlibrary{quantikz}
\usetikzlibrary{calc}

\begin{document}


\title{Formation of Tesseract Time Crystals on a Quantum Computer}


\author{Christopher Sims}
\affiliation{Elmore Family school of Electrical and Computer Engineering, Purdue University}
\email{Sims58@purdue.edu}


\date{\today}

\begin{abstract}
The engineering of new states of matter through Floquet driving has revolutionized the field of condensed matter physics. This technique enables the creation of hybrid topological states and ordered phases that are absent in normal systems. Crystalline structures, exemplifying spatially ordered systems under periodic driving, have been extensively studied. However, the recent focus has shifted towards discrete time crystals (DTCs), periodically driven quantum many-body systems that break time translation symmetry under specific conditions. In this paper, we explore the theoretical predictions, experimental realizations, and emerging possibilities of utilizing DTCs on quantum computers. Additionally, the formation of time varying tesseracts using discrete time crystals is presented, allowing for the investigation of time translational symmetry in higher-dimensional lattice systems.
\end{abstract}

\maketitle

\section{Introduction}
The engineering of new states of matter using Floquet driving has revolutionized the field of condensed matter physics \cite{Wang2013,Bukov2015,Harper2020,Titum2016}. By employing this technique, researchers have successfully created hybrid topological states and ordered phases in time that were previously unattainable. While periodic driving of crystalline structures has been extensively investigated, the emergence of discrete time crystals (DTCs) has opened up new avenues for exploration \cite{GomezLeon2013}. Unlike their spatial counterparts, DTCs exhibit periodicity in discrete time steps rather than in spatial dimensions.

The theoretical foundation of DTCs is rooted in the concept of periodically driven quantum many-body systems that break time translation symmetry \cite{Else2016,Keyserlingk2016,Sacha2017,Wilczek2012,Watanabe2015,Huse2013,Pekker2014}. These theoretical predictions have led to the identification of distinctive phases, including the strongly disordered thermal phase and the many-body localization phase\cite{Ge2022,Koshkaki2022}. Subsequent experimental investigations have successfully confirmed the existence of DTCs in various systems, such as cold atoms and condensed matter systems driven by periodic laser pulses \cite{Basko2006,Zhang2017,Choi2017,Rovny2018}.

The exploration of DTCs has extended beyond one-dimensional systems, captivating the interest of researchers in higher-dimensional lattice structures \cite{Kuros2021}. These higher-dimensional systems offer additional degrees of freedom for studying time translational symmetry and the occurrence of multiple distinct phases within a single system \cite{Kuros2021}. Recent experimental breakthroughs have observed the formation of two-dimensional discrete time crystals in crystal systems, shedding light on the dynamics and properties of DTCs in higher dimensions \cite{Gao2022,Yoshida2015,Lee2022}. It has been shown what when DTCs are extended to higher dimensions they become more resilient to noise and thermal disorder \cite{Pizzi2021,Pizzi2021a}

Quantum computers have emerged as powerful tools for studying and manipulating DTCs. These cutting-edge machines enable the simulation of complex quantum systems, facilitating investigations into the behavior of DTCs in larger systems. Quantum computers also provide insights into the role of quantum entanglement and allow for the exploration of various parameters and driving protocols that influence the stability and properties of DTCs. Furthermore, the unique control capabilities of quantum computers make it possible to design optimal driving protocols for creating and sustaining DTCs \cite{Mi2021,Frey2022}. Advanced control techniques, such as optimal control theory and machine learning algorithms, can be employed to devise robust time-evolution protocols. Recently, it has been shown that higher dimensional time crystals can be formed on a quantum computer by taking advantage of hardware connectivity \cite{Sims2023}.

To further advance the understanding of time translational symmetry in higher-dimensional systems, a novel concept has emerged: the formation of tesseracts using discrete time crystals. Tesseracts represent four-dimensional analogs of cubes and their formation using DTCs presents an intriguing opportunity. Exploring tesseract formations allows for the investigation of the interplay between time translational symmetry and higher-dimensional dynamics. Moreover, it provides valuable insights into the stability, robustness, and emergent phenomena associated with DTCs in dimensions $N\geq2$ \cite{Pizzi2021,Pizzi2021a}.

This work presents the study of the formation of temporal tesseracts on a quantum computer by leveraging orthogonal rotations and MBL interactions from quantum algorithm. This work demonstrates that it is possible to form higher dimensional objects beyond 2D on a quantum computer by leveraging both time translational symmetry and spatial translational symmetry.

\section{Discrete Time Crystals}
1D DTC can be described as a many-body Bloch-Floquet hamiltonian with phase flip and disorder terms \cite{Nandkishore2015,Abanin2019,Ponte2015,Lazarides2015,Bordia2017}.
\begin{equation}
\begin{aligned}
H = &  {} H_1  + H_2 + H_3\\
H_1 =  & g(1-\epsilon) \sum_i \sigma_i^y \\
H_2 =  & \sum_i J_{<i,j>} \sigma_i^x \sigma_j^x \\
H_3 =  & \sum_i D_{i} \sigma_i^x 
\end{aligned}
\label{BFH}
\end{equation}
Where $J_{<i,j>}$ is the Ising coupling,  $D_{i}$ is the disordered term, and $g(1-\epsilon)$ is the rotation term \cite{Throckmorton2021,Alet2018,Chen2022,Hu2017,Kjaell2014,Hauke2015}.

In order to form DTCs on a quantum computer a seperate algorithm needs to be implemented, the following shows the floquet unitary for a 1D DTC system:
\begin{equation}
U_F =  \prod_{j=1}^{N-1} \exp^{-i\theta  X_j}  \prod_{j=1}^{N-1} \exp^{-i\phi Z_j Z_{j+1}}  \prod_{j=1}^{N-1} \exp^{-i\varphi XZ}
\label{UF}
\end{equation}
Where the gate angles are $\theta = h \tau_1$ ($h = g(1-\epsilon)$), $\phi = J(\tau_2 - \tau_1)$, and $\varphi = D(T-\tau_2)$. Where the final wavefunction is $\ket{\psi_n} = U_F^n \ket{\psi_0}$, where $\ket{\psi_0}$ is the initial state. While this unitary is an example, any orthogonal set of gates can be defined and utilized to form a DTC on a quantum computer.

In this work the phased ZZ gate is defined as:
\begin{equation}
ZZ(\phi) = 
\begin{bmatrix}
1 & 0 & 0 & 0 \\
0 &  \exp^{i\phi} & 0 & 0 \\
0 & 0 & \exp^{i\phi} & 0 \\
0 & 0 & 0 & 1 \\
\end{bmatrix}
\end{equation}

\section{A 4D Discrete Time Crystal}
Previous works cover the formation of a 1D and a 2D discrete time crystal on a quantum computer. However, in order to extend to higher time dimensions as well, the orthogonal rotations need to be taken advantage of for a 4D rotation. The 4D point system can be diffing in two spatial components and two time components $q  = [x , y , \theta, \phi]$. since X and Y correspond to the qubit locations on a quantum computer. A rotation in the qubit space is considered. As long as the rotations are kept to be orthogonal, it is possible to construct a pseudo rotation about an axis for a 4D object.

The 4D DTC modeled with the Cirq quantum simulator for 50 floquet time steps with 16 qubits. $\theta = \pi/2$ and $\phi = \pi/2$ are the terms chosen because they are the only terms with lead to a symmetric rotation about an the 4$^{\text{th}}$ axis. This is because all rotations are in the quantum system are isometric, as oppose to the true rotation matrix which allows for any arbitrary rotation where the relation between all points is conserved. 

\begin{figure}[!ht]
\centering
 \includegraphics[width=1.0\textwidth]{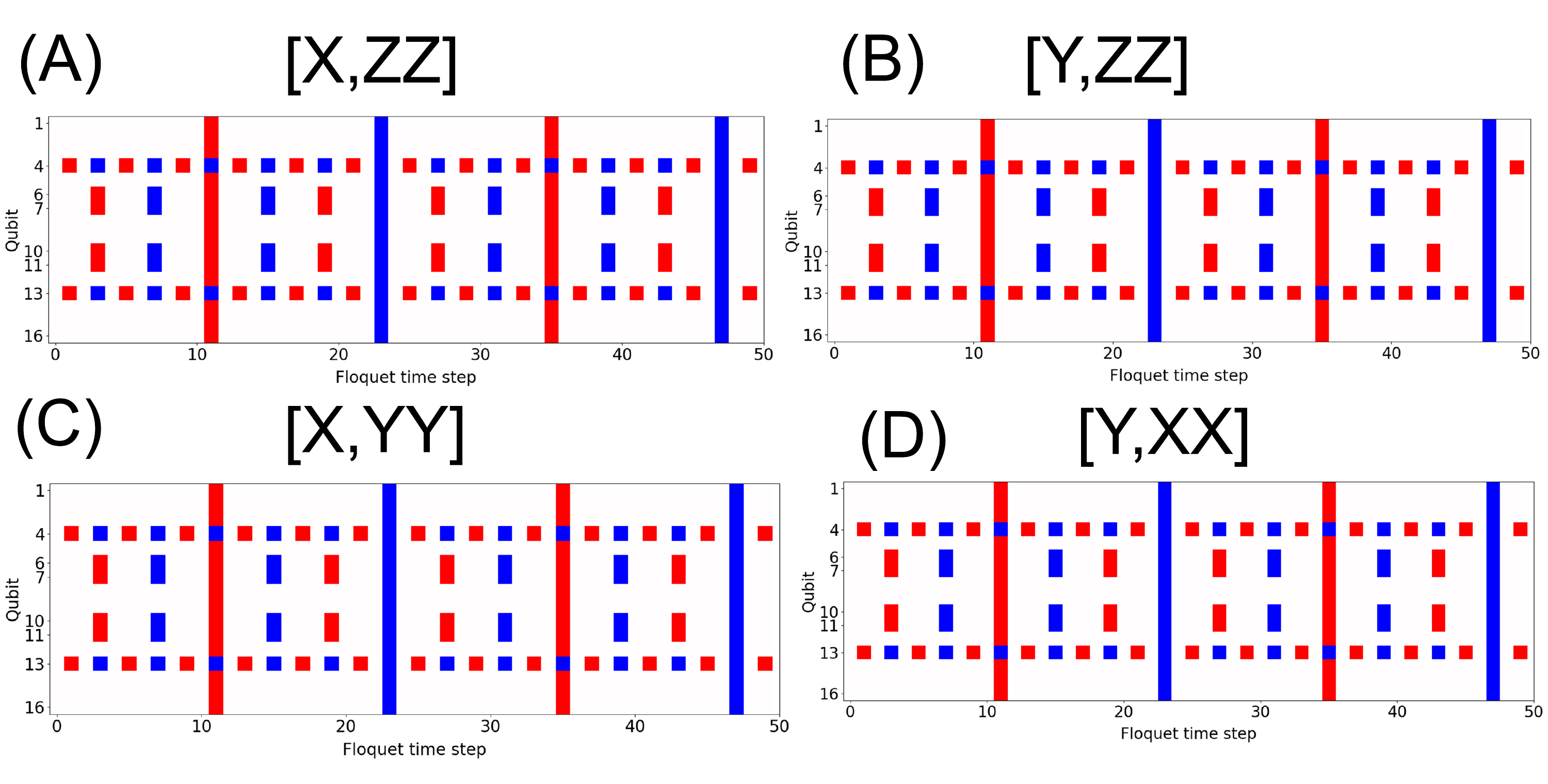}%
 \caption{\textbf{Polarization measurement:} The polarization measurement $\overline{A}$ = $\overline{\langle \ket{1}Z_i(t)\rangle}$ of each qubit for each time step $t$ in the DTC state with $\theta = \pi/2$ and $\phi = \pi/2$ (A) Rotation in X and interaction in ZZ (B) Rotation in Y and interaction in ZZ (C) Rotation in X and interaction in YY (D) Rotation in Y and interaction in XX. All measurements are made in the  $\hat{Z}$ basis. Red = $\ket{1}$,blue $\ket{0}$}
\label{polZ}
 \end{figure}

When the qubits are rotated with the phased X gate and the interaction term is orthogonal with the ZZ gate. It is possible to realize a 4D object with a rotation about the 4$^{\text{th}}$ axis. This corresponds to a rotation of a tesseract in 4D. This $\pi/2$ rotation means that a plane in 2D is crossed every 4 iterations $2\pi$, and in order for the system to rotate to its initial state, a total of 24 $\pi/2$ rotations are required. The system starts out with the initial state $\ket{\psi_0}_i = \ket{0}$. The first iteration rotates all qubits $\pi/2$ to the X axis. In the absence of the interaction term, the next rotation will bring all qubits to the $\ket{1}$ state, however, only the qubits on the corners $q_6,q_{13}$ are in the $\ket{1}$ state. This shows that the interaction terms are creating a higher dimensional object with pseudo rotations about a higher dimensional axis. All even interactions do not intersect a higher dimensional plane, so do not show any states in the computational basis. The 3rd iteration shows that a box has formed in the inside of the 2D qubit grid ($q_6,q_7,q_{10},q_{11}$). This inner box, corresponds to the ``inner box'' of the hypercube that is a tesseract. After $2\pi$ rotations, the other side of the inner cube appears in the 2D qubit grid. After another set of $2\pi$ rotations, the outer cube plane shows as a fully $\ket{1}$ filled qubit grid. This process repeats throughout the Unitary gates which shows that this is indeed a 4D hypercube which has formed in a quantum circuit [Figure \ref{polZ}(A)] . The corner modes in qubits 4 and 13 arise due to edge interactions from the discrete edge that forms due to real hardware limitations, if the qubits are arranged with no edge conditions, these edge modes disappear. Figure \ref{polZ}, shows the unitary with different sets of orthogonal rotation gates. Figure \ref{polZ}(B), rotate in Y and interaction terms in ZZ. Figure \ref{polZ}(C), rotate in X and interaction terms in YY. Figure \ref{polZ}(D), rotate in Y and interaction terms in XX.
	
\begin{figure}[!ht]
\centering
 \includegraphics[width=1.0\textwidth]{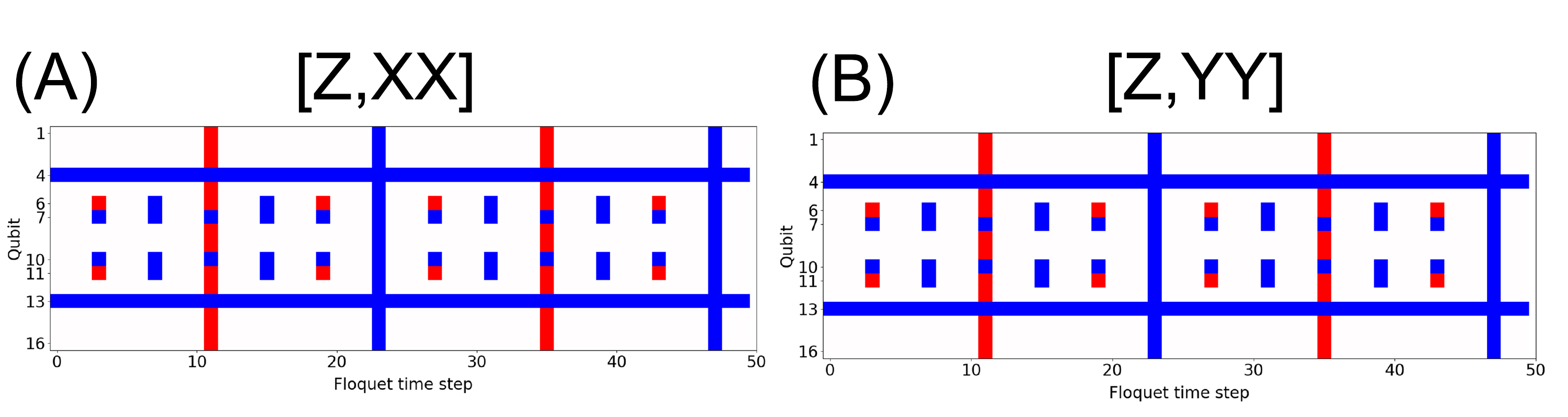}%
 \caption{\textbf{Polarization measurement:} The polarization measurement $\overline{A}$ = $\overline{\langle \ket{1}Z_i(t)\rangle}$ of each qubit for each time step $t$ in the DTC state with $\theta = \pi/2$ and $\phi = \pi/2$ for rotations in the Z axis with (A) XX interactions and (B) ZZ interaction. Red = $\ket{1}$, blue $\ket{0}$.}
\label{polZZ}
 \end{figure}

	Within the computational basis, a rotation in the Z axis with interaction terms in its orthogonal terms corresponds to an equivalent rotation about the 3$^{\text{rd}}$ axis for a higher dimensional geometric object. With this rotation, one edge of the hypercube is in view with a plane crossing the 2D projected plane every full rotation (24 time steps) [Figure \ref{polZZ}]. The mixed state on some of the planes are due to edge interactions which cause a phase shift. Orthogonal interaction gates XX and YY show the same results [Figure \ref{polZZ}(A,B)].
	
	The 4D discrete time crystal is an expanded form of the 2D discrete time crystal. In addition to the qubits being arranged in a ``virtual'' 2D crystal which corresponds to the 2D grid layout of the quantum computer, the rotation gates are forced to be orthogonal and the disorder term is eliminated from the unitary. 
In order to demonstrate the 2D nature of the qubits and their time evolution, the qubit structure is rearranged into a 2D grid and a time evolution is applied to the unitary operations. From the 2D animation it is easier to see the 4D rotations with a 2D projection $A = P_{x,y}R_{\theta}$, where $P_{x,y}$ corresponds to projection onto a 2D axis [Figure \ref{2DpolZ}].

\begin{figure}[!ht]
\centering
\animategraphics[width=0.5\linewidth, autoplay, loop,]{4}{X_ZZ_}{0}{49}
 \caption{\textbf{2D Polarization measurement:} The polarization measurement $\overline{A}$ = $\overline{\langle \ket{1}Z_i(t)\rangle}$ of each qubit for each time step $t$ in the DTC state with $\theta = \pi/2$ and $\phi = \pi/2$ for rotations in the X axis and ZZ interactions Red = $\ket{1}$, blue $\ket{0}$.}
\label{2DpolZ}
 \end{figure}

\newpage
\section{Discussion}
Higher spatial dimension time crystals have more connectivity than 1D time crystals. The Ising interaction $J_{i,j}$ increased the interaction terms for the coupling between two qubits. The Ising interaction terms can be considered as an `edge' to a higher dimensional object, and qubits correspond to edges with their own respective phases. This representation can be utilized as a graph representation of a mathematical object. Hypercubes, or tesseracts, have 8 edge connectivity for each vertex, however, sometimes the edges are aligned and only show 4 edge connectivity. By providing 8 interaction terms on a 2D lattice and 2 time orthogonal rotations, a 2D projection of a 4D time dimensional (2D + 2) object is formed. With this, a 16 vertex (qubit) object forms a tesseract in the discrete time-crystal system is formed. Such an object can be used as timing on a quantum computer, or as a way to control novel algorithms on quantum computer. 

\section{Methods}
\subsection{Simulation}
All calculations were performed in python v3.9.1 with the Google quantum AI Cirq package v1.0.0. GPU acceleration is enabled with the Nvidia cuQuantum 22.07.1 SDK.  Qubits were simulated with a grid layout similar to Google quantum machines. All qubits are typically measured in the $\hat{Z}$ basis in a quantum machine. However, for this simulation, qubits are measured using CIRQ's vector state simulator, the polarization measurement is then done in the $\hat{Z}$ basis in order to get the qubit rotation with respect to the $\hat{Z}$ axis. When mentioned in the text, the qubits are also measured in the $\hat{X}$ basis. The resulting measurement is then extracted with the autocorrelator.
\subsection{Edge State}
When computing the average of the qubit states, all qubits are taken into consideration. Edge qubits and their effects are considered to be part of the system and are integral to the results.
\subsection{Circuits}
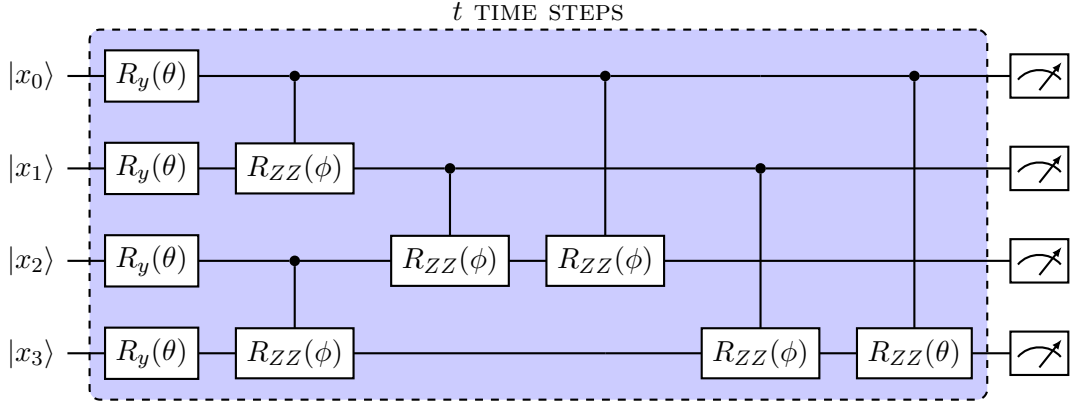
\begin{figure}[!h]
\centering
\begin{quantikz}
\lstick{\ket{x_0}} & \gate{R_y(\theta)} \gategroup[4,steps=6,style={dashed,
rounded corners,fill=blue!20, inner xsep=2pt},
background]{{\sc $t$ time steps}} & \ctrl{1} &\qw&\ctrl{2}&\qw&\ctrl{3} &\meter{}\\
\lstick{\ket{x_1}} & \gate{R_y(\theta)} &  \gate{R_{ZZ}(\phi)} & \ctrl{1} &\qw&\ctrl{2}&\qw &\meter{}\\
\lstick{\ket{x_2}} & \gate{R_y(\theta)}  &\ctrl{1} &  \gate{R_{ZZ}(\phi)}&\gate{R_{ZZ}(\phi)}& \qw&\qw  &\meter{}\\
\lstick{\ket{x_3}} & \gate{R_y(\theta)}  & \gate{R_{ZZ}(\phi)} & \qw&\qw&\gate{R_{ZZ}(\phi)}&\gate{R_{ZZ}(\theta)}&\meter{}
\end{quantikz}
\caption{\textbf{2D DTC circuit}: An example of a 2D DTC crystal algorithm with 4 qubits implemented, this results in a 2$\times$2 square with interacting qubits. This algorithm is repeated for $t$ Bloch-Floquet time steps and then measured at the end. $\vec{x}$ corresponds to the initial state, which can be of N\'eel, polarized, or random order.}
\label{circ}
\end{figure}
The time crystal algorithm comprises three main stages: rotations, mixing parameters, and long-range interactions, denoted as $H_1$, $H_2$, and $H_3$ respectively (as shown in Equation \ref{BFH}). Translating the Bloch-Floquet time step into a quantum algorithm involves three simple gates. The first stage involves single-gate rotations, specifically $R_y$ gates. In this work, all gates are rotated by $\pi$ as it is the most stable and optimal rotation for a real quantum algorithm.

The next stage involves the "spin" interactions, which can be ideally modeled using a two-qubit $R_{ZZ}$ gate. Although the more generalized FSIM gates have been utilized in previous works to fine-tune the circuit, this work focuses on the $R_{ZZ}$ variation. To form a two-dimensional state, the $R_{ZZ}$ gates are applied to the nearest neighbors (NN) with qubits arranged in a theoretically square pattern. Notably, since Google quantum computers are arranged in a square pattern, this algorithm can be effectively implemented using an NN algorithm.

\section{Conclusion}
In conclusion, a method is developed to form tesseracts in a quantum computer by utilized a 2D grid of qubits with orthogonal spin rotations. This work shows that is is possible to construct $N\geq 2$ dimensional time crystals by leveraging orthogonal rotations about an arbitrary time axis'. In addition, this work demonstrates the first experimental realization of a higher dimensional mathematical object. In future work, gates can be constructed in the unitary so that new pseudoplanes can be constructed in qubit space and new orthogonal rotations can be added to a higher dimensional spatial and temporal time crystal (e.g q = $[ x, y, \theta_1,\theta_2....,\theta_i] $).
\section{Code availability}
Code will be made publicly available upon full publication under the GNU GPL v3.0 license.

\section{Supplementary information}
See supplementary information for equivalent geometric analogues, more detailed information about gates, and a more detailed view of the autocorrelators.

 \begin{acknowledgments}
C.S. acknowledges the generous support from the GEM Fellowship and the Purdue Engineering ASIRE Fellowship.
 
 Correspondence and requests for materials should be addressed to C.S.
 (Email: Sims58@Purdue.edu)
\end{acknowledgments}


%

\end{document}